# Strain Induced One-Dimensional Landau-Level Quantization in Corrugated Graphene

Lan Meng[1], Wen-Yu He[1], Hong Zheng[1], Mengxi Liu[2], Hui Yan[1], Wei Yan[1], Zhao-Dong Chu[1], Keke Bai[1], Rui-Fen Dou[1], Yanfeng Zhang[2,3], Zhongfan Liu[2], Jia-Cai Nie[1], and Lin He[1,*]

[1] Department of Physics, Beijing Normal University, Beijing, 100875, People's Republic of China
[2] Center for Nanochemistry (CNC), College of Chemistry and Molecular Engineering, Peking University, Beijing 100871, People's Republic of China
[3] Department of Materials Science and Engineering, College of Engineering, Peking University, Beijing 100871, People's Republic of China

Theoretical research has predicted that ripple of graphene generates effective gauge field on its low energy electronic structure and could lead to Landau quantization. Here we demonstrate, using a combination of scanning tunneling microscopy and tight-binding approximation, that Landau levels will form when effective pseudomagnetic flux per ripple $\Phi \sim (h^2/la)\Phi_0$ is larger than the flux quantum $\Phi_0$ (here $h$ is the height, $l$ is the width of the ripple, $a$ is the nearest C-C bond length). The strain induced gauge field in the ripple only results in one-dimensional (1D) Landau-level quantization along the ripple. Such 1D Landau quantization does not exist in two-dimensional systems in an external magnetic field. Its existence offers a unique opportunity to realize novel electronic properties in strained graphene.

## I. INTRODUCTION

The charge carriers in graphene display a linear dispersion resulting from the geometrical structure [1-7]. This linear dispersion ensures that the low-energy behaviors of the charge carriers can be described by the two-dimensional massless Dirac equation. Lattice deformations, which change the electron hopping between sublattices, give rise to perturbations in the Dirac equation. The change in the hopping between sublattices can be described mathematically as effective gauge fields, which affect the Dirac fermions in graphene like an effective magnetic field [8-12]. Experimental observation of Landau level-like quantization in strained graphene demonstrated that the deformation-induced gauge field is an experimental reality [13-16]. Therefore, it's expected to realize zero-field quantum Hall effect in strained graphene. Theoretically, the pseudo-Landau levels in graphene were first introduced in relation to the rippling-induced partially flat bands at zero-energy [9] and have been further confirmed by *ab initio* calculations [10]. It was believed that the ripples, which can be either intrinsic or induced by a substrate, have important consequences affecting the electronic properties in graphene, such as electron localization and charge inhomogeneity [9,10,17-22]. However, a direct evidence for the emergence of the Landau levels quantization in graphene ripples is still lacking so far.

In this paper, we demonstrate, using a combination of scanning tunneling microscopy and tight-binding approximation, that the Landau levels will form when the effective pseudomagnetic flux per ripple $\Phi \sim (h^2/la)\Phi_0$ is larger than the flux quantum $\Phi_0$ (here $h$ is the height, $l$ is the width of the ripple, $a$ is the distance between neighboring carbon atoms). More importantly, the lattice deformation of the ripple only results in one-dimensional (1D) Landau-level quantization along the ripple but not in the perpendicular direction. Such 1D Landau-level quantization is distinct from that observed previously in the irregular strained graphene systems [13,14] and is impossible to be realized in two-dimensional systems in an external magnetic field. Its existence could offer a unique opportunity to realize novel electronic properties in strained graphene. The condition to generate the ripples with $h^2/la \geq 1$ in the corrugated graphene is also discussed according to classical thin-film elasticity theory.

## II. EXPERIMENTAL METHODS

The graphene monolayer was grown on a 25 micron thin polycrystalline Rh foil, via a traditional ambient pressure chemical vapor deposition (CVD) method, as reported in previous papers [19,23,24]. The as-grown sample was synthesized at 1000 ºC and then cooled down to low temperature for scanning tunneling microscopy (STM) characterization. Owing to the thermal expansion mismatch between the graphene and the Rh foil, defect-like wrinkles and ripples tend to evolve along the boundaries of crystalline terraces for strain relief [19,24-26]. In a previous paper, we demonstrated that the quasi-periodic ripples with small curvature in the corrugated graphene give rise to weak 1D electronic potentials and lead to the emergence of the superlattice Dirac points [19]. In this paper we focus on the structure and electronic properties of ripples with large curvature. The STM system was an ultrahigh vacuum four-probe scanning probe microscope from UNISOKU. All the STM and scanning tunneling spectroscopy (STS) measurements were performed in an ultrahigh vacuum chamber ($10^{-10}$ Torr) and all the images were taken in a constant-current scanning mode at liquid-nitrogen temperature. The STM tips were obtained by chemical etching from a wire of

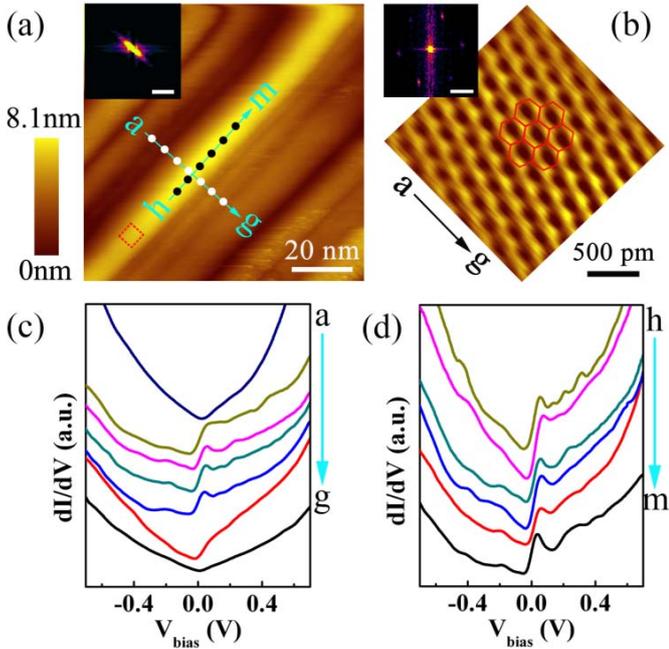

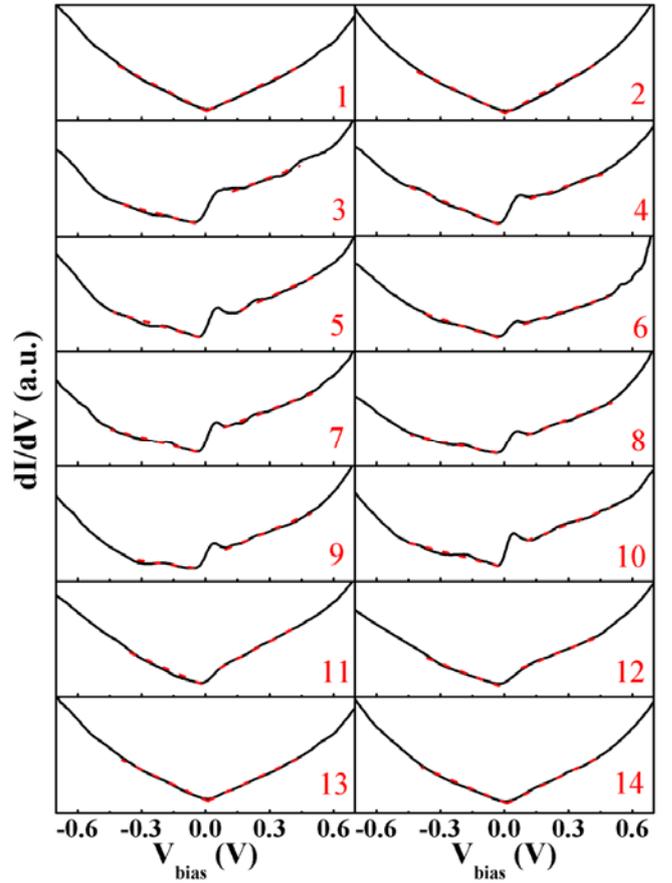

**FIG. 1** (color online). (a) A STM image of corrugated graphene with 1D ripples showing a wide distribution of periods on Rh foil ($V_{sample}$ = 1.20 V and I = 0.21 nA). The inset is Fourier transform of the main panel showing the reciprocal-lattice of the ripples. The scale bar is 1 nm$^{-1}$. (b) A zoom-in topography in the red frame of panel (a) shows the observed triangular pattern on the ripple ($V_{sample}$ = -330 mV and I = 0.14 nA). The atomic structure of graphene is overlaid onto the STM image. The inset is Fourier transform of panel (b). The scale bar is 20 nm$^{-1}$. (c) Tunneling spectra recorded at different positions, as marked by the dots along the arrow from a to g, in panel (a). The spectra were vertically offset for clarity. (d) The dI/dV-V curves, recorded at different positions, as marked by the dots along the arrow from h to m, in panel (a). A pronounced peak close to the Fermi energy is observed in the tunneling spectra of the ripple.

**FIG. 2** (color online). Fourteen tunneling curves from top to down are recorded at positions a (curves 1 and 2), b (curves 3 and 4), c (curves 5 and 6), d (curves 7 and 8), e (curves 9 and 10), f (curves 11 and 12), and g (curves 13 and 14), respectively. The red dashed lines flanking the Dirac points are used to show the slop of these tunneling spectra.

Pt(80%) Ir(20%) alloys. Lateral dimensions observed in the STM images were calibrated using a standard graphene lattice. The tunneling spectrum, *i.e.*, the dI/dV-V curve, was carried out with a standard lock-in technique using a 789 Hz alternating current modulation of the bias voltage.

### III. EXPERIMENTAL RESULTS

Figure 1(a) shows several ripples of a typical corrugated graphene on Rh foil. The ripples show a wide distribution of periods and thereby the reciprocal-lattice of these ripples displays a wide distribution of values, as shown in the inset of Fig. 1(a). Usually, the amplitude of these ripples is very small and the local curvature of the ripples does not break the six-fold symmetry of the graphene lattice. Then the honeycomb lattice of graphene, including both the A and B sublattices, can be observed by STM measurements [19]. A corrugated graphene with a large curvature (or strain) is expected to break the symmetry between the A and B sublattices and result in the so-called "three for six" triangular pattern of atoms [18]. One of these ripples in Fig. 1(a) shows a large curvature that can break the symmetry of the sublattices. This is exactly shown in Fig. 1(b) in which a clear triangular lattice (either the A or the B sublattice) was observed on the graphene ripple. The width and the amplitude of the ripple are estimated to be 21-28 nm and 3.0-3.5 nm respectively. On the assumption that the ripple can be approximated as a sinusoid, then the strain roughly estimated from measuring the surface length of the graphene ripple yields value of 0.18 [27]. This large local strain not only breaks the symmetry of the sublattices, but also changes the topological features of the ripple's electronic states.

Figure 1(c) shows seven dI/dV-V curves recorded at different positions from a to g along the arrow in Fig. 1(a). The tunneling spectrum gives direct access to the local density of states (DOS) of the surface at the position of the STM tip. Curves a, f, and g, which are measured at ripples with small curvature, show a linear DOS around the Fermi level. This is consistent well with the tunneling spectrum of graphene monolayer [19]. The minimum dip at about ~ 0 V is attributed to the Dirac point of graphene. Curves b, c, d, and e, which show distinct characteristics comparing with that of the curves a, f, and g, are measured at the ripple with a large strain. The experimental result in Fig. 1(c) indicates that the ripple with a large strain shows distinct electronic band structure from that of the ripples with a small curvature.

For the ripple with a large strain, a pronounced peak near the Fermi energy is clearly resolved in the spectrum. The intensity of the tunneling peak is almost independent of the positions along the ripple, as shown in Fig. 1(d). The tunneling peak of the ripple is attributed to the strain-induced partially flat bands at zero-energy [9,10]. The slight departure of the peak from the Fermi energy, ~ 0.04 eV, may arise from the charge transfer between the substrate and the corrugated graphene. Similar tunneling peak, which also originates from the zero-energy flat bands, was observed in the tunneling spectrum of topological line defects of graphite/graphene [28]. Besides the zero-energy peak, the spectra of the ripple show two other features. The first one is the asymmetry of the slope of the tunneling curve flanking zero-bias, as shown in Fig. 2. The slopes of the tunneling spectra in the positive bias ($S_R$) and negative bias ($S_L$) show a clear asymmetry. This behavior was mainly attributed to the electron-hole asymmetry, which originates from next-nearest-neighbor hopping $t'$. Owing to the lattice deformation, the next-nearest-neighbor hopping is expected to be enhanced and thereby it is not surprised to observe the large electron-hole asymmetry in the strained graphene structures [13,14].

## IV. ANALYSIS AND DISCUSSION

To further understand the effect of ripples on the tunneling spectra, we investigate the electronic band structure and DOS of the 1D graphene ripples by tight-binding approximation. In our model, we assume that the height field of graphene ripples can be approximated by a sinusoidal function $H(x) = h\sin\left(\dfrac{2\pi x}{l}\right)$ [9,10,29], as shown in Fig. 3(a). Here $h$ is the amplitude and $l$ is the period of the ripples. The arc length infinitesimal of $H(x)$ can be expressed as $ds = \sqrt{[H'(x)]^2 + 1}\,dx$. If the condition that $H'(x) \ll 1$ holds (which requires $h/l \ll 1$), the differential of the arc length can be approximated as $ds = \left\{1 + \dfrac{1}{2}[H'(x)]^2\right\}dx$. Then the local deformation along the $x$ axis can be written as $ds - dx = \dfrac{1}{2}[H'(x)]^2 dx$. For simplicity, the strain tensor associated with the local deformation along the $x$ axis can be written as $u_x = 1/2[H'(x)]^2$ and the deformation along the $y$ axis is negligible [9,10]. In weakly deformed lattice, the nearest-neighbor hopping parameter is changed to [11]

$$t_i = t + \dfrac{\beta t}{a^2}\vec{\rho}_i\cdot(\vec{u}_i - \vec{u}_0).$$

Here $\beta = -\dfrac{\partial \ln t}{\partial \ln a} \approx 2$ is the Grüneisen parameter, $a$ is the lattice constant, $\vec{\rho}_i$ ($i$=1, 2, 3) are the nearest-neighbor vectors and $\vec{u}_i - \vec{u}_0$ is substituted by

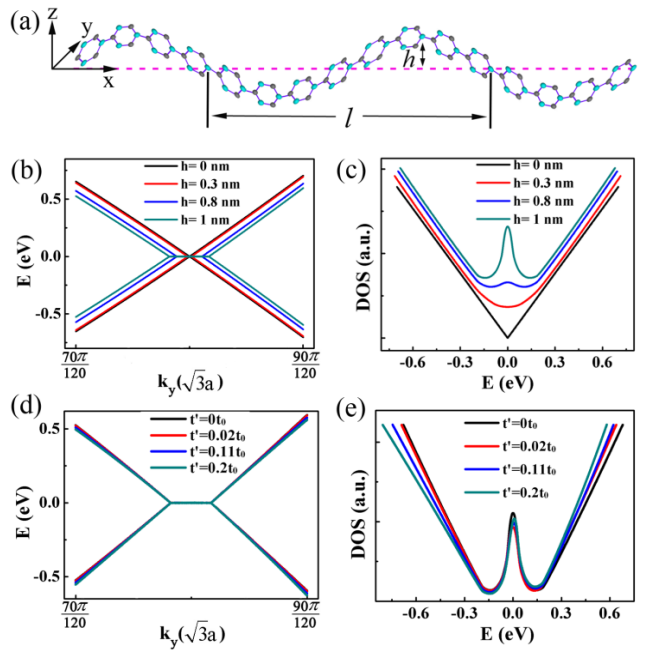

**FIG. 3** (color online). (a) Perspective view of two graphene ripples, which can be approximated as a sinusoid. The period and amplitude of the ripples are $l$ and $h$ respectively. (b) The lowest energy band of a graphene ripple with $(h, l)$ = (0, 21), (0.3, 21), (0.8, 21), and (1, 21), respectively (both $h$ and $l$ are in unit of nm). In the calculation, we only take into account the nearest-neighbor hopping $t_0$ and set the next nearest-neighbor hopping $t'$ = 0. (c) The corresponding local DOS of the ripples in panel (b). (d) The lowest energy band of a graphene ripple with $(h, l)$ = (1, 21) and $(t_0, t')$ = (3 eV, 0 eV), (3 eV, 0.06 eV), (3 eV, 0.33 eV), and (3 eV, 0.6 eV), respectively. For clarity, the energy of all the band structure is shifted $3t'$ to make the flat bands stay at zero energy. For the case $t' \neq 0$, the electronic band structure of the ripple is not symmetry about the zero energy. (e) The corresponding DOS of the ripples in panel (d). For clarity, the energy of all the DOS is shifted $3t'$ to make the peak of the DOS stays at zero energy. When $t' \neq 0$, the slope of the DOS flanking the zero-energy is not symmetry and the asymmetry increases with increasing the $t'$.

$(\vec{\rho}_i\cdot\nabla)\vec{u}(\vec{r})$ according to the continuum limit (elasticity theory). For our one-dimensional graphene ripples with $l$ = 150$a$ (100 unit cells), a 200×200 matrix can be established. Through diagonalizing the matrix of Hamiltonian $H$, we can calculate the energy spectrum of the ripples. The density of states of the ripples is calculated numerically according to the well-known formula $\dfrac{S}{4\pi^2}\oint\dfrac{1}{|\nabla_k E(\vec{k})|}dk$.

In our calculation, we fixed the period of the ripples as about 21 nm ($l$ = 150$a$) and changed the amplitude of the ripples. The electronic band structure and the DOS of the 1D graphene ripples are shown in Fig. 3(b) and Fig. 3(c), respectively. Obviously, the ripple with a large curvature (or strain) can really generate the flat bands and consequently the apparent DOS peak at zero-energy. Both the range of $k_y$ values, where the flat bands are well defined, and the intensity of the DOS peak increase with increasing the amplitude of the ripples. As mentioned in

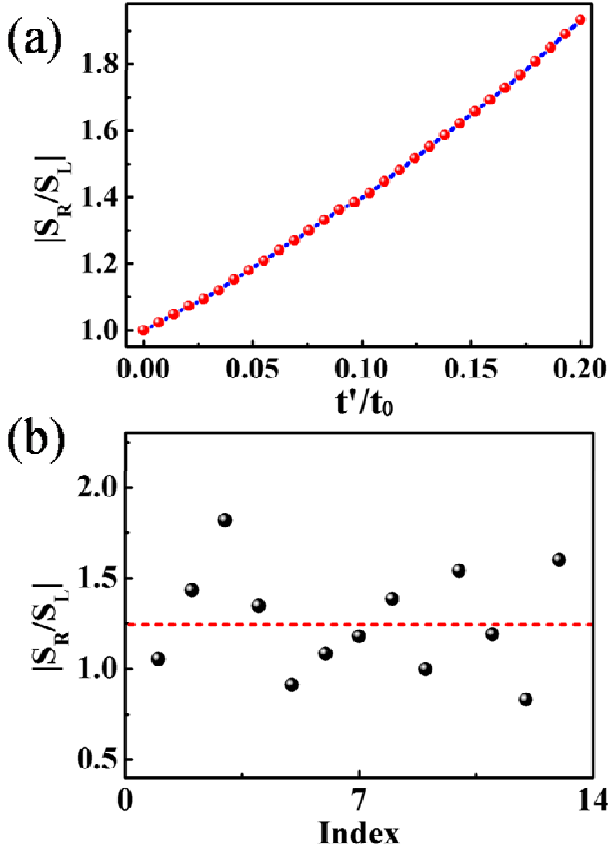

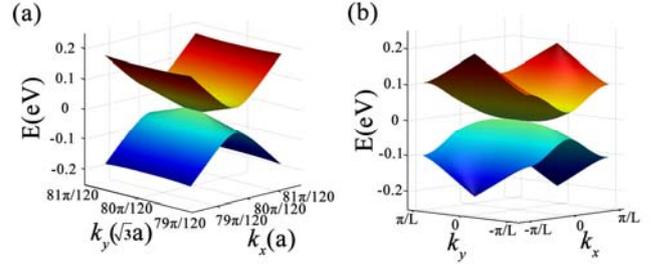

FIG. 5 (color online). (a) Energy dispersions of the graphene ripples. (b) Energy dispersions of the 1D graphene superlattice. Both band structures show anisotropy of the Fermi velocity.

**FIG. 4** (color online). (a) The ratio of the slope $|S_R/S_L|$ in the calculated density of states as a function of $t'/t_0$. (b) The ratio of the slope $|S_R/S_L|$ of the tunneling spectra obtained in our experiment. The dotted line shows the average ratio $|S_R/S_L| \sim 1.25$.

the preceding chapter, the rippling-induced hopping modulation in graphene can be described as an effective gauge field generated in real magnetic fields. The lattice deformation of the ripples results in partially flat bands at zero-energy, as shown in Fig. 3(b), which are the analog of Landau levels.

When considering the next nearest-neighbor hopping t', the tight-binding Hamiltonian for the sinusoidal graphene ripples becomes

$$H = -\sum_{m,i}\left[t_i(x)a_m^\dagger b_{m+\rho_i} + H.c.\right]$$

$$-t'\sum_{m,n,j}\left(a_m^\dagger a_{m+\Delta_j} + b_n^\dagger b_{n+\Delta_j} + H.c.\right),$$

where $a_m$ and $b_n$ are the annihilation term on sublattices A and B, respectively. The next nearest-neighbor hopping t' breaks the electron-hole symmetry of the ripples in the corrugated graphene, as shown in Fig. 3(d) and 3(e). The asymmetry of the DOS around zero energy increases with increasing the value of t', as shown in Fig. 4(a). Therefore, our analysis indicates that the ripples can result in zero-energy flat bands and a large electron-hole asymmetry in the corrugated graphene, which are both in agreement with our experimental results, as shown in Fig. 4(b).

Here we should point out that the rippling-induced gauge field only results in zero-energy flat bands in the parallel direction of the ripples (the $k_y$ direction) but not in another. The Fermi velocities in the perpendicular direction of the ripples (the $k_x$ direction) are not renormalized at all, as shown in Fig. 5(a). It reveals that the lattice deformation in the 1D ripples leads to anisotropy of the Fermi velocities in the corrugated graphene. Similar anisotropy of the Fermi velocity could also be caused in the pristine graphene by a 1D periodic potential, as shown in Fig. 5(b). The absence of Fermi velocity reduction in the $k_x$ direction is closely related to the chiral tunneling of electrons in graphene incident in the normal direction of a potential barrier [19,30].

The second feature of the spectra shown in Fig. 1(d) is the emergence of other weak peaks at high bias. Figure 6(a) shows a typical tunneling spectrum of the ripple. The intensity of the peaks at high bias is much weaker than that at zero-energy. This agrees with the fact that the higher pseudo-Landau levels are less well defined and the description of a hopping modulation as an effective gauge field is exact valid only at the Dirac point [9]. Although weak, five peaks can still be observed in the spectrum. The peak positions $E_n$ (= $eV_{bias}$) follow the expression [13-16]

$$E_n = E_D + \text{sgn}(n)v_F\sqrt{2e\hbar B_S|n|}, \quad n = 0, \pm 1, \pm 2,..., \quad (1)$$

quite well, as shown in Fig. 6(b). Here, $v_F = 1.0 \times 10^6$ m/s is the Fermi velocity of monolayer graphene, $\hbar$ is the Planck's constant, $B_S$ is the pseudomagnetic field, and $E_D$ is the position of the Dirac point. A reasonable linear fit yields $B_S \sim 49$ T. This behavior can be understood as that

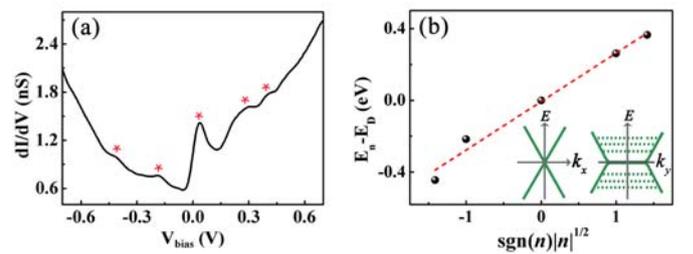

FIG. 6 (color online). (a) A typical dI/dV-V curve (dark curve) of the graphene ripple (this curve is reproduced from that of Fig. 1(d)). The red asterisks denote the positions of Landau levels. (b) The energy of pseudo-Landau level peaks $E_n$-$E_D$ deduced from the spectrum in panel (a) as a function of $\text{sgn}(n)|n|^{1/2}$. The red dashed line is the linear fit. The inset shows schematic energy band structure of the ripple, which shows linear energy dispersion in the $k_x$ direction and Landau level quantization in the $k_y$ direction.

the rippling-induced several partially flat bands in the band structure of graphene at discrete energies that can be roughly estimated by Eq. (1) with a fitting parameter $B_S$.

Another intrinsic effect, which weakens the signal of the Landau levels in the tunneling spectra, is that the discrete Landau levels of the ripple are generated only in the $k_y$ direction and the energy dispersion is linear and continuous along the $k_x$ direction, as shown in the inset of Fig. 6(b). In the irregular strained graphene systems [13,14], the strain induced Landau quantization in both the $k_x$ and $k_y$ directions, which mimics the case that an external magnetic field applied perpendicular to the sample. Therefore, the 1D Landau-level quantization is distinct from that observed previously in strained graphene [13,14] and more importantly this unique system is impossible to be realized in two-dimensional systems in an external magnetic field.

Theoretically, the gauge field induced by the strain can be written as [11]

$$A_x = -2\frac{\beta}{a}u_{xy}, \quad A_y = \frac{\beta}{a}(u_{xx} - u_{yy}).$$

Here, the strain tensor $u_{ij}(\mathbf{r})$ is associated to the deformation of the graphene ripples by [31]

$$u_{xx} = \frac{\partial u_x}{\partial x} + \frac{1}{2}\left(\frac{\partial H}{\partial x}\right)^2,$$

$$u_{yy} = \frac{\partial u_y}{\partial y} + \frac{1}{2}\left(\frac{\partial H}{\partial y}\right)^2,$$

$$u_{xy} = \frac{1}{2}\left(\frac{\partial u_x}{\partial y} + \frac{\partial u_y}{\partial x}\right) + \frac{1}{2}\frac{\partial H}{\partial x}\frac{\partial H}{\partial y}.$$

Owing to the fact that the corrugation of the ripple is along $x$ axis, only $u_{xx}$ is nonzero. Then the pseudomagnetic field in the ripple can be obtained as

$$B_s \sim -\frac{\beta}{a}\frac{4\pi^3 h^2}{l^3}\sin\frac{4\pi x}{l}.$$

In our experiment, we did not observe the space-dependent pseudomagnetic field described by the above expression. This may mainly arise from the fact that the structure of the ripple deviates much from a sinusoidal function. This periodic pseudo-magnetic field along $x$ axis localizes a pseudo-Landau level wave function in an area of $l^2/16$. So the magnetic flux corresponding to this area is $\Phi \sim \frac{h^2}{la}$ [9]. The zero-energy states caused by inhomogeneous magnetic field is observable under the condition of $\Phi > 1$, for the reason that the perpendicular magnetic field with total N flux quantum leads to (N-1) zero-energy states [32]. Therefore, it was predicted that the Landau levels will form in graphene ripples when the effective magnetic length of the pseudomagnetic field is smaller than the ripple size, i.e., $h^2/la \geq 1$, and for the case that $h^2/la < 1$, the electronic band structures not differ much from that of flat graphene monolayer [9]. Our experimental result and analysis, as shown subsequently, demonstrate that the condition $h^2/la \geq 1$ is very difficult to satisfy.

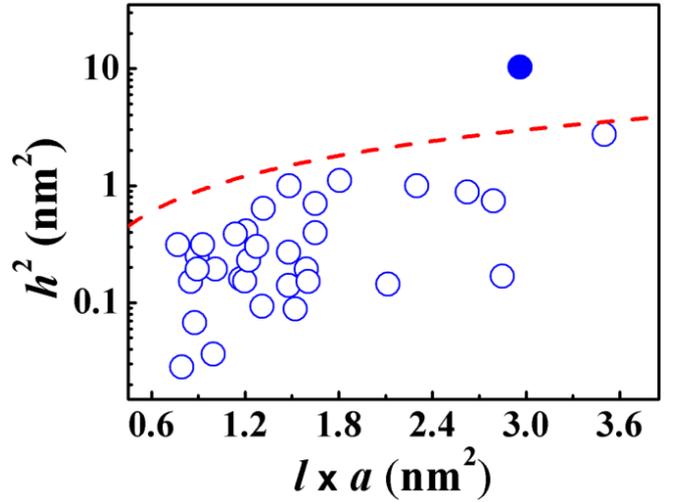

**FIG. 7** (color online). This figure summarizes the parameters, $l$ and $h$, of the ripples studied in our experiment. The filled circle represents the ripple showing Landau levels and the open circles are denoted the ripples that not show any evidence of the Landau level. The red dashed curve is plotted according to $h^2/la = 1$.

Figure 7 summarizes the parameters, $l$ and $h$, of more than thirty ripples studied in our experiment. Only one of them satisfies $h^2/la \geq 1$ and thereby the Landau levels are observed. For the ripples with $h^2/la < 1$, the obtained tunneling spectra are almost identical to that of flat graphene monolayer. Usually, the ripple formation process in corrugated graphene can be understood based on the classical thin-film elasticity theory [33,34]. For simplicity, we assume that the ripples are generated in the presence of a longitudinal tensile strain $\gamma$. Then, according to classical elasticity theory, the amplitude $h$ and the period $l$ can be obtained as [33]

$$l = \left[\frac{4\pi^2(tL)^2}{3(1-v^2)\gamma}\right]^{1/4},$$

$$h^2 = vtL\left[\frac{16\gamma}{3\pi^2(1-v^2)}\right]^{1/2}.$$

Here $v$ is the Poisson ratio, $t$ is the thickness of graphene, and $L$ is the length of ripple. Consequently, we can obtain $\frac{h^2}{la} = \frac{v}{a}\left[\frac{64\gamma^3 t^2}{3\pi^6(1-v^2)}\right]^{1/4}\sqrt{L}$, where $v$ is the Poisson ratio, $t$ is the thickness of graphene, and $L$ is the length of the ripples [34]. For the case of a small strain, for example $\gamma = 0.015$, the condition $h^2/la \geq 1$ is satisfied when the length of the ripple $L \geq 8$ μm. Experimentally, the length of the ripple is observed to be much shorter than 8 μm. Therefore, it's not very easy to observe the zero-energy Landau levels in graphene ripples with a small strain. For the ripple with $\gamma = 0.18$, the condition $h^2/la \geq 1$ is satisfied when $L \geq 0.2$ μm. We should further point out that the above analysis is a very rough estimation. The effects of the surface structure of the substrate, the contribution of other type of strain, and the deviation of the ripples from a sinusoidal function were not taken into account in the model. Further experiments by using other substrates or by using substrate with pre-define structures maybe helpful to induce the necessary strained ripples.

## V. CONCLUSIONS

In summary, we demonstrated that the ripples, which satisfy $h^2/la \geq 1$, can result in Landau-level quantization and a large electron-hole asymmetry in the corrugated graphene. Our analysis further reveals that the lattice deformation in the 1D ripples leads to the Landau levels quantization in one direction but not in the other. This unique system provides a platform to explore novel electronic properties of strained graphene in the near future.

## ACKNOWLEDGMENTS

We acknowledge helpful discussions with Yugui Yao. This work was supported by the National Natural Science Foundation of China (Grant No. 11004010, No. 10974019, No. 21073003, No. 51172029 and No. 91121012), the Fundamental Research Funds for the Central Universities, and the Ministry of Science and Technology of China (Grants No. 2011CB921903, No. 2012CB921404, No 2013CB921701). L.M. and W.-Y.H. contributed equally to this paper.

*Email:helin@bnu.edu.cn.